# Time Reversal Induces Negative Mass and Charge Conjugation: On the Physical Interpretation of the Irreducible Unitary Representations of Negative Mass and Energy of the Full Poincare Group

J.N. Pecina-Cruz
Department of Physics and Geology
The University of Texas-Pan American
Edinburg, Texas 78501-2999
jpecina2@panam.edu

**Abstract**

This article proves that according to the principles of quantum mechanics the existence of elementary particles of negative mass is physically plausible. Heisenberg's uncertainty principle plays an important role in this demonstration. By reversing time, charge and mass are conjugated and the symmetry group is that of the Poincare with space-time reflections (full Poincare group). Particles that tunnel from the future time-like cone to the past time-like cone move backward in time with negative mass (and energy). These particles are interpreted as antiparticles of positive mass (and energy) that move forward in time. There is not a physical law that precludes two observers from seeing different mass signs of a particle, since the mass is not a Poincare invariant, but its square $p^2 = -m^2$.

**Introduction**

A popular misunderstanding in physics states the mass of an elementary particle is a Casimir invariant of the Poincare group. The group of Poincare has two Casimir, which are invariants under the transformations of the group. The eigenvalue for one of these Casimir is the mass square, but not the mass. This means that the sign of the mass of an elementary particle could be seen different for two observers. Therefore, according to this argument nothing prevents to have particles with negative mass [4]. With the interpretation of particles moving backward in time as antiparticles, one can argue that the current account of absorption and emission of particles is inconsistent with a mass balance. This inconsistence disappears if particles moving backward in time have negative mass (and energy). This argument is discussed in section 2. Section 1 presents the interpretation of antiparticles according to reference 2. Section 3 is devoted to the construction of the irreducible unitary representations of the full Poincare group (Poincare group with simultaneous space-time reflection.) And to demonstrate that time reversal, naturally, induces negative mass and charge conjugation.

## 1. Antiparticles

In the one particle scheme Feynman and Stückelberg interpret antiparticles as particles moving backwards in time [3]. This argument is reinforced by S. Weinberg who realizes that the antiparticles existence is a consequence of the violation of the principle of causality in quantum mechanics [2]. The temporal order of the events is distorted when a particle wanders in the neighborhood of the light cone. How is the antimatter generated from matter? According to Heisenberg's uncertainty principle, a particle wandering in the neighborhood of the light-cone suddenly tunnels from the timelike region to the spacelike; in this region the relation of cause and effect collapses. Since if an event, at $x_2$ is observed by an observer A, to occur later than one at $x_1$, in other words $x_2^0 > x_1^0$. An observer B moving with a velocity **v** respect to observer A, will see the events separated by a time interval given by

$$x_2'^0 - x_1'^0 = L_\alpha^0(v)(x_2^\alpha - x_1^\alpha), \tag{1}$$

where $L_\alpha^\beta(v)$ is a Lorentz boost. From equation (1), it is found that if the order of the events is exchanged for the observer B, that is, $x_2'^0 < x_1'^0$ (the event at $x_1$ is observed later than the event at $x_2$.), then a particle that is emitted at $x_1$ and absorbed at $x_2$ as observed by A, it is observed by B as if it were absorbed at $x_2$ before the particle were emitted at $x_1$. The temporal order of the particle is inverted. This event is completely feasible in the neighborhood of the light-cone, since the uncertainty principle allows a particle to tunnel from time-like to space-like cone regions. That is the uncertainty principle will consent to the space-like region reach values above than zero as is shown in next equation,

$$(x_1 - x_2)^2 - (x_1^0 - x_2^0)^2 \leq \left(\frac{\hbar}{mc}\right)^2,$$

where $\left(\frac{\hbar}{mc}\right)^2 > 0,$ (2)

and $\frac{\hbar}{mc}$ is the Compton wave-length of the particle. The left hand side of the equation (2) can be positive or space-like for distances less or equal than the square of the Compton wavelength of the particle. Therefore, causality is violated. The only way of interpreting this phenomenon is assuming that the particle absorbed at $x_2$, before it is emitted at $x_1$, as it is observed by B, is actually a particle with negative mass and energy, moving backward in time; that is $t_2 < t_1$ [2]. This event is equivalent to see an antiparticle moving forward in time with positive mass and energy, and opposite charge and spin (as it is discussed in the next section) that it is emitted at $x_1$ and it is absorbed at $x_2$. With this reinterpretation the causality is recovered.

The square of the mass for the observers A and B is a Poincare invariant. In other words, *the mass of a particle itself is not an invariant, but its square* [4]. There is nothing to prevent that the rest mass could have different sign in distinct reference frame, as happen with its energy, electric charge and spin. In this paper, it is conjectured that a particle moving backward in time possesses a negative mass. When it is observed as an antiparticle moving forward in time its mass is positive. Still its mass square is a Poincare invariant.

This section is concluded with the remark that the transition from positive mass to negative mass is an uphill event, since a particle has to overpass the speed of light. Also, this violation of the classical laws of physics is only temporary and regulated by the uncertainty principle. Therefore, the arguments against the existence of negative mass are unattainable.

**2 Absorption of a Particle and Emission of an Antiparticle**

Figure 1 shows a particle of mass m that it is moving backward in time and it is absorbed by a target. The same physical event as it is seen by a second observer is shown in Fig. 2. According to the current interpretation [2] (positive mass of the particle that it moves backward in time,) the target increases its mass by m. Figure 2 shows an antiparticle that moves forward in time, as it is seen by second observer. The antiparticle is emitted by the target. Therefore, the target decreases its mass by –m. Hence, the current interpretation of a particle moving backward in time by an antiparticle moving forward in time is inconsistent with a mass balance. A way to circumvent this problem is to assume that the mass of a particle absorbed by the target is negative. The mass could change sign for two different observers (since $p^2 = -m^2$). Another solution would be to take the mass of the antiparticle as negative, leaving in the target a hole of positive mass. Both assumptions are physically interesting and have immediate experimental consequence.

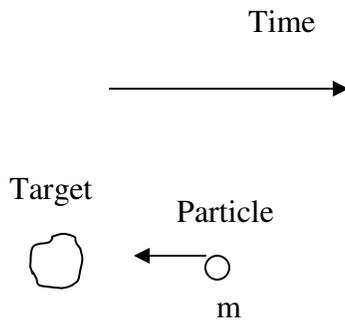

**Figure 1.** The target gains a mass m

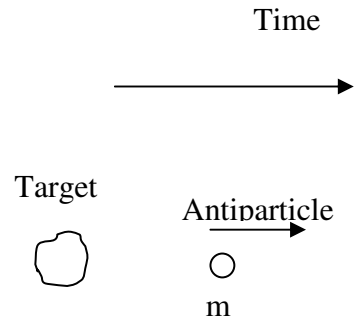

**Figure 2.** The target loses a mass m

The disintegration of a neutron inside of a nucleus can lead to the emission of a positron and the transformation of a neutron into a proton. This kind of disintegration decreases the mass of the nucleus and rules out the second possibility.

## 3. Representations of Negative Mass

Let us choose the vectors $|\hat{k}\rangle$ as the basis vectors of an irreducible representation $T^{(k)}$, of the subgroup of translations of the Poincare group. Then the basis vectors $|\hat{k}'\rangle$, with $\hat{k}' = W\hat{k}$, and $W$ a Lorentz transformation, lies in the same representation. That is, $k'$ and $k$ are in the same region of the space-time; these vectors have the same "length", $(k',k') = (k,k)$.

$$T(\hat{u})T(W)|\hat{k}\rangle = T(W)|\hat{k}\rangle \exp(W\hat{k},\hat{u}) \tag{3}$$

The vectors $T(W)|\hat{k}\rangle$ transform according to the irreducible representation $T^{(Wk)}$, of the subgroup of translations [5].

Let us start our analysis with the future time-like cone or region 2. By choosing the time-like four-vector $\hat{k}_0 = (0,0,0,k)$ with $k>0$, it is found that the little group corresponding to the orbit of the point $(0,0,0,k)$ is the group of rotations in three dimensions SU(2).

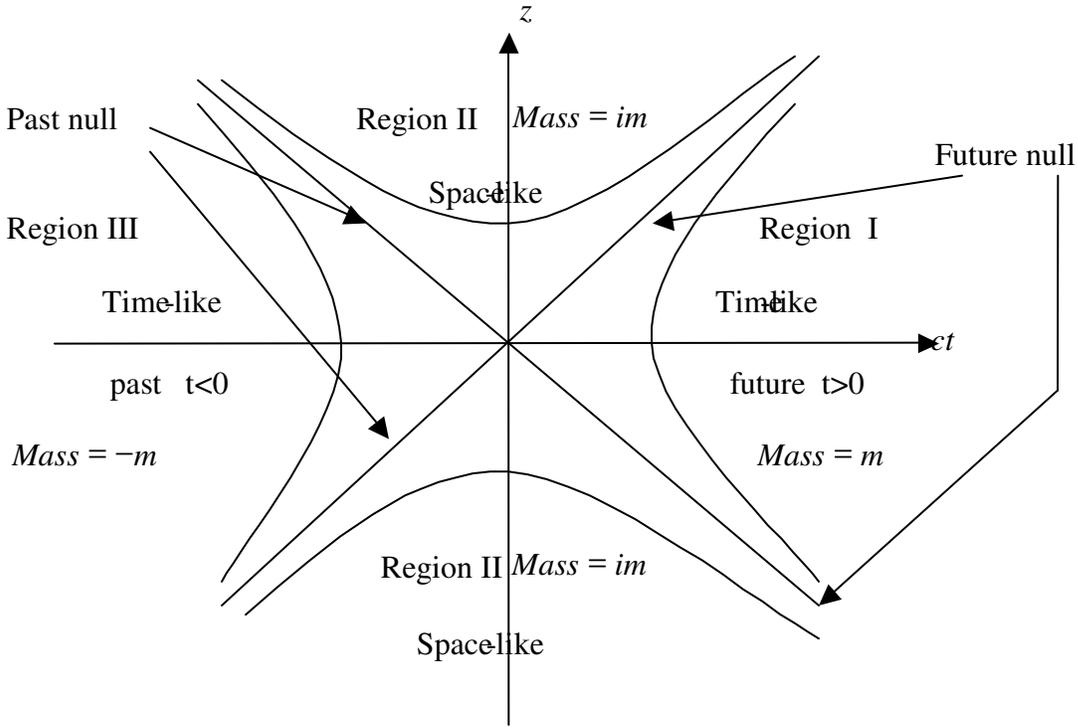

**Fig. 3. Light-cone**

Therefore, the representation is labeled by $\hat{k}_0$ and by an irreducible representation label of the three dimensional rotational group SU(2). Starting from the 2s + 1 basis vectors of the little group SU(2) the bases vectors $\left|\hat{k}sm_s\right\rangle$, of an irreducible representation of the Poincare group, are generated by a pure Lorentz transformation (a boost) which carries $\hat{k}_0$ into $\hat{k}$, that is $W_k \hat{k}_0 = \hat{k}$. The basis vectors of the representation are given by

$$\left|\hat{k}sm_s\right\rangle = T(W_k)\left|\hat{k}_0 sm_s\right\rangle. \tag{4}$$

This group operation preserves the "length" of the vector $\hat{k}_0$, that is $(k,k) = (k_0, k_0)$. By applying a Lorentz transformation followed by a translation to equation (4) can be proved that the vectors $\left|\hat{k}sm_s\right\rangle$ furnish a unitary irreducible representation of the Poincare group

$$T(\hat{u})T(W)\left|\hat{k}sm_s\right\rangle = \left|\hat{k}'sm_s'\right\rangle D_{m_s'm_s}(W_{k_0}) \exp(\hat{k}', \hat{u}). \tag{5}$$

Where $\hat{k}' = W\hat{k}$, and $(W\hat{k}, W\hat{k}) = (\hat{k}, \hat{k})$. $D_{m_s'm}(W_{k_0})$ are the unitary irreducible representations of the little group SU(2), $W$ is an arbitrary Lorentz transformation. The representations are unitary because the generators of the group are unitary, and irreducible because the bases vectors of the representations are generated from a single vector $\left|\hat{k}_0 sm_s\right\rangle$, with linear momenta equal to zero. Their equivalent unitary irreducible representations are denoted by $P^{(k_0,s)}$. These representations have the same length of the four-vector $\hat{k}$ than that of $\hat{k}_0$. The set of inequivalent representations with different magnitude of the four-vector $\hat{k}$ are denoted by $P^{(k,s)}$.

Let us construct the unitary irreducible representations of the Poincare group with simultaneous space-time reflections. First note that if the vectors $\left|\hat{k}\right\rangle$ are the bases of an irreducible representation of the translation group, and $I$ is the space-time reflection operator, then

$$T(\hat{u})T(I)\left|\hat{k}\right\rangle = T(I)T(I\hat{u})\left|\hat{k}\right\rangle = T(I)\left|\hat{k}\right\rangle \exp(I\hat{k}, \hat{u}) \tag{6}$$

Therefore, the vectors $T(I)\left|\hat{k}\right\rangle$ transform according to the $T^{(Ik)}$ representation of the translation subgroup of the Poincare group. But, the operator $I$ reverse the space and time components of $\hat{k}$, that is, the energy $E = \hbar c k_t$ becomes negative. On the rest frame of the $P^{(k,s)}$ representation, the rest mass, $m_0 = \dfrac{\hbar I k_0}{c} = -\dfrac{\hbar k_0}{c}$, of such a particle would also be negative. There is nothing that prevents this happen, since $m_0^2$ is a Poincare invariant [4], but not $m_0$. That is, if one observer sees a particle with rest mass positive other observer on an inertial frame could see the same particle with negative rest mass. An electron with negative rest mass and energy moving backward in space and time could be interpreted as its antiparticle moving forward in space and time with positive rest mass and energy. This phenomenon is explained by constructing the unitary irreducible representations of the Poincare group with simultaneous space-time inversions.

To construct the unitary irreducible representations of the extended Poincare group one applies the space inversion followed by time inversion to the basis vectors $\left| \hat{k} s m_s \right\rangle$ of $P^{(k,s)}$. These basis vectors were generated from the special basis vector $\left| \hat{k}_0 s m_s \right\rangle$, where $\hat{k}_0 = (0,0,0,k)$. The space inversion $I_s$ leaves $\hat{k}_0$ invariant, and commutes with the generators of the little group SU(2). Therefore, the irreducible representations of the group, $Z_2 \times SU(2)$, can be labeled by the labels of the rotations, and reflection groups, namely the spin s, and the parity η. If one applies space inversions to the basis vector $\left| \hat{k}_0 s m_s \eta \right\rangle$ with definite parity label η = ±, one obtains

$$T(I_s)\left| \hat{k}_0 s m_s \eta \right\rangle = \left| \hat{k}_0 s m_s \eta \right\rangle \eta \tag{7}$$

On the rest frame the parity η, is an eigenvalue of the basis vector $\left| \hat{k}_0 s m_s \eta \right\rangle$. Now, Let us start to construct the inequivalent unitary irreducible representations of the Poincare group with space inversions $P^{(k,s,\eta)}$. By applying is to the basis vectors $\left| \hat{k} s m_s \eta \right\rangle$, given by equation (4), and including the parity label η one obtains

$$T(I_s)\left| \hat{k} s m_s \eta \right\rangle = T(I_s)T(W_k)\left| \hat{k}_0 s m_s \eta \right\rangle = T(W_{-k})T(I_s)\left| \hat{k}_0 s m_s \eta \right\rangle = $$
$$T(W_{-k})\left| \hat{k}_0 s m_s \eta \right\rangle \eta = \left| I_s \hat{k} s m_s \eta \right\rangle \eta \tag{8}$$

The Lorentz boost $W_{-k}$ changes the direction of space components of $\hat{k}$. Equation (8) shows that the basis vector $\left| \hat{k} s m_s \eta \right\rangle$ is an invariant space under space reflections; the action of the space inversion operator on a general basis vector leads to another basis vector. Hence, these bases vectors yield a representation for the Poincare group with space inversions.

If one applies linear time reflections to the basis vector $\left| \hat{k}_0 s m_s \eta \right\rangle$, one finds that

$$T(I_t)\left| \hat{k}_0 s m_s \eta \right\rangle = \left| I_t \hat{k}_0 s m_s \eta \right\rangle \tag{9}$$

Since, from equation (6) $T(I_t)|\hat{k}_0 s m_s \eta\rangle$ transforms according to the $T^{(I_t k_0 s)}$ representation of the subgroup of translations. That is, the basis vector $|\hat{k}_0 s m_s \eta\rangle$ with $\hat{k}_0 = (0,0,0,k)$ from the $P^{(k_0,s,\eta)+}$ representation is taken into the basis vector $|I_t \hat{k}_0 s m_s \eta\rangle$ with $I_t \hat{k}_0 = (0,0,0,-k)$ of the $P^{(k_0,s,\eta)-}$ representation [5]. We will show that in the $P^{(k_0,s,\eta)-}$ representation time reflections prevent that SU(2) commutes with $Z_2$. In spite of SU(2) does not commute with $Z_2$, in the $P^{(k_0,s,\eta)-}$ representation, still η is a label for the full Poincare group representation. Since the Casimir operators of the full Poincare groups are $P^2$, $W^2$, and $I^2$.

If the vector $|\hat{k}\rangle$ is any vector that transforms according to a representation of the group of translations, then

$$\hat{P}|\hat{k}\rangle = -i\hat{k}|\hat{k}\rangle \tag{10}$$

Hence, in the $P^{(k_0,s,\eta)+}$

$$\hat{P}|\hat{k}_0 s m_s \eta\rangle = -ik|\hat{k}_0 s m_s \eta\rangle \tag{11}$$

While in the $P^{(k_0,s,\eta)-}$ representation

$$\hat{P}|I_t \hat{k}_0 s m_s \eta\rangle = +ik|I_t \hat{k}_0 s m_s \eta\rangle \tag{12}$$

Then, time reflection induces a change on the energy sign, that is

$$I_t P_0 I_t^{-1} = -P_0 \tag{13}$$

That is, time inversions do not commute with $P_0$.

The Pauli-Lubanki four vector components on the rest frame in the $P^{(k_0,s,\eta)+}$ representation is

$$W_q |\hat{k}_0 s m_s \eta\rangle = -k J_q |\hat{k}_0 s m_s \eta\rangle, \text{ and}$$

$$W_t = 0, q = x, y, z \tag{14}$$

Now, in the $P^{(k_0,s,\eta)-}$ representation, on the orbit of the vector $I_t \hat{k}_0 = (0,0,0,-k)$

$$W_q \left| I_t \hat{k}_0 s m_s \eta \right\rangle = +k J_q \left| I_t \hat{k}_0 s m_s \eta \right\rangle, and$$

$$W_t = 0, q = x, y, z$$

(15)

So that, time inversions do not commute with the Pauli-Lubansky four-vector in the larger space composed by $P^{(k_0,s,\eta)+}$ and $P^{(k_0,s,\eta)-}$.

$$I_t W_q I_t^{-1} = -W_q,$$

(16)

Therefore, from equation (14) one obtains

$$I_t J_q I_t^{-1} = -J_q$$

(17)

That is, time reversal induces an inversion on the direction of a rotation and changes the sign of the rest energy (rest mass) of the particle. Then, by using equation (17) one gets

$$J_z I_t \left| \hat{k}_0 s m_s \eta \right\rangle = -I_t J_z \left| \hat{k}_0 s m_s \eta \right\rangle = I_t \left| \hat{k}_0 s m_s \eta \right\rangle (-m_s)$$

(18)

The vector $I_t \left| \hat{k}_0 s m_s \eta \right\rangle$ transforms like $-m_s$, under rotations about the z-axis. Therefore from equation (17) for a general rotation, we get

$$T(R(\theta))T(I_t) \left| \hat{k}_0 s m_s \eta \right\rangle = T(I_t)T(R(-\theta)) \left| \hat{k}_0 s m_s \eta \right\rangle =$$

$$T(I_t) \sum_{m_s'} \left| \hat{k}_0 s m_s' \eta \right\rangle D^{(s)-1}_{m_s' m_s}(\theta) = T(I_t) \sum_{m_s'} \left| \hat{k}_0 s m_s' \eta \right\rangle D^{(s)*}_{m_s' m_s}(\theta)$$

(19)

Hence, the vectors $T(I_t) \left| \hat{k}_0 s m_s \eta \right\rangle$ transform like the transpose conjugate complex, $D^{(s)*}_{m_s' m_s}$, of the representation $D^{(s)}_{m_s' m_s}$ of the unitary irreducible representation of SU(2). This fact explains why it is necessary to conjugate and transpose the wave equation of a particle to describe its antiparticle. Thus, time reflection induces negative energy states and these states induce charge conjugation. If a mirror reflection is a symmetry transformation, then this reflection must be accompanied by simultaneous space-time inversions, since the intrinsic parity label is generated by space reflection.

Due to the fact that one has to conjugate and transpose, time reversal acquires the properties of an antilinear operator. And since the character of the representations of the three-dimensional rotation group is real, the representations, $D^{(s)*}_{m'_s m_s}$ and $D^{(s)}_{m'_s m_s}$, are equivalent and those representations can be reached one from the other by a change of basis. We apply this transform to the transpose of the transpose of complex conjugated representation. On the new basis the basis vector is given by

$$\left| \hat{k}_0 s m_s \eta \right\rangle_{\substack{new \\ basis}} = \sum_{m_s} \delta^{m'_s}{}_{-m_s} (-)^{s-m'_s} \left| \hat{k}_0 s m'_s \eta \right\rangle \tag{20}$$

Since the matrix $D^{(s)}$ transforms into the matrix $D^{(s)*}$ according to a set of similarity transformations. As a matter of fact we are mapping a vector over its dual. Then, one has to define

$$D^{(s)*}_{m'_s m_s} = F^{-1} D^{(s)}_{m'_s m_s} F^1 \tag{21}$$

where the transformation matrix $F$ is given by

$$F = \delta^{m'_s}{}_{-m_s} (-)^{s-m_s} \tag{22}$$

By applying the linear time reversal operator, on the new basis vector, to the general basis vector $\left| k s m_s \eta \right\rangle$ which, is defined by equation (4), one gets

$$T(I_t) \left| \hat{k} s m_s \eta \right\rangle = \sum_{m'_s} T(I_t) \left| \hat{k} s m'_s \eta_s \right\rangle \delta^{m'_s}{}_{-m_s} (-)^{s-m'_s} = \left| I_t \hat{k} s - m_s \eta \right\rangle (-)^{s+m_s} \tag{23}$$

Under the action of the time reversal operator a particle on the rest frame reverses its energy, and spin.

We define the scalar product according to reference [5] by

$$\left| \psi \right\rangle = \sum_{m_s} \int \psi_{m_s}(\hat{k}) \left| \hat{k} s m_s \right\rangle k_t^{-1} dk \tag{24}$$

Therefore applying the bra-vector one gets

$$\psi'_{m_s}(\hat{k}) = (-1)^{s-m_s} \psi^*_{-m_s}(I_t \hat{k}) \tag{25}$$

Here, the conjugation is a consequence of the bra-vector. Applying the space reversal operator to the general basis vector $\left| k s m_s \eta \right\rangle$, one obtains

$$T(I_s)|\hat{k}sm_s\eta\rangle = T(W_{-k})T(I_s)|\hat{k}_0 sm\eta_s\rangle = T(W_{-k})|\hat{k}_0 sm_s\eta\rangle\eta = |I_s\hat{k}sm\eta_s\rangle\eta \qquad (26)$$

The momentum is reversed and the particle acquires a definite parity. In order to generate the basis vectors of the representation for the Poincare group with simultaneous space-time reflections, we apply the time reversal operator to equation (8)

$$T(I)|\hat{k}sm_s\eta\rangle = T(I_t)T(I_s)T(W_k)|\hat{k}_0 sm_s\eta\rangle = T(I_t)T(W_{-k})|\hat{k}_0 sm_s\eta\rangle\eta$$
$$= T(I_t)|I_s\hat{k}sm_s\eta\rangle\eta = |I\hat{k}s - m_s\eta\rangle\eta(-)^{s+m_s} \qquad (27)$$

Therefore, in this formulation the simultaneous action of space-time inversions on a general basis vector (particle) of the representation reverses the energy, momentum, and spin (antiparticle). The space inversion furnishes a definite parity, η, to the elementary particle described by the unitary irreducible representation. The elementary particle violates the causality principle because it is represented in the negative energy sector of the light cone. If it has enough energy can be observed as a particle moving backwards in space and time (antiparticle). Therefore, the C, T, P, CT, CP, PT cannot conserve separately, but CPT. In my opinion no all the symmetries that could enhance the full Poincare group are known. Then, it is quite probable a possible violation of CPT for physical phenomena that requires more large symmetries. The full group of diffeomorphisms (space-time reflections included) is a larger symmetry than that of the full Poincare group; therefore CPT could be violated by quantum gravity. By the same token supersymmetry, supergravity, and superstrings could violate CPT.

From equation (27) we get

$$T^2(I)|\hat{k}sm_s\eta\rangle = |\hat{k}sm_s\eta\rangle(-)^{2s} \qquad (28)$$

Hence, $T^2 = 1$ for integer spin, and $T^2 = -1$ for particles of half integer spin.

**Conclusion**

The acceptance of negative mass in physics is a controversial matter. However, there are traces of its possible existence. This paper presents a logic scheme, where the concept of negative mass is reasonable according to the principles of quantum mechanics.